\documentclass[prb,showpacs,preprintnumbers,amsmath,amssymb,twocolumn]{revtex4-1}
\usepackage{graphicx}
\usepackage{dcolumn}
\usepackage{bm}
\usepackage{epstopdf}
\usepackage{natbib}
\usepackage{color}
\begin{document}

\title{Possible devil's staircase in the Kondo lattice CeSbSe}

\author{K. -W. Chen$^{1,2}$, Y. Lai$^{1,2}$,  Y. -C. Chiu$^{1,2}$, S. Steven$^3$, T. Besara$^1$, D. Graf$^{1}$, T. Siegrist$^{1,4}$, T. E. Albrecht-Schmitt$^3$, L. Balicas$^{1,2}$, and R. E. Baumbach$^{1,2}$
}
\affiliation{$^1$National High Magnetic Field Laboratory, Florida State University}
\affiliation{$^2$Department of Physics, Florida State University}
\affiliation{$^3$Department of Chemistry and Biochemistry, Florida State University}
\affiliation{$^4$Department of Chemical and Biomedical Engineering, FAMU-FSU College of Engineering}
\date{\today}

\begin{abstract}
The temperature ($T$) - magnetic field ($H$) phase diagram for the tetragonal layered compound CeSbSe, is determined from magnetization, specific heat, and electrical resistivity measurements. This system exhibits complex magnetic ordering at $T_{\rm{M}}$ $=$ 3 K and the application of a magnetic field results in a cascade of magnetically ordered states for $H$ $\lesssim$ 1.8 T which are characterized by fractional integer size steps: i.e., a possible Devil's staircase is observed. Electrical transport measurements show a weak temperature dependence and large residual resistivity which suggest a small charge carrier density and strong scattering from the $f$-moments. These features reveal Kondo lattice behavior where the $f$-moments are incompletely screened, resulting in a fine balanced magnetic interaction between different Ce neighbors that is mediated by the RKKY interaction. This produces the nearly degenerate magnetically ordered states that are accessed under an applied magnetic field.
\end{abstract}

\maketitle

\section{Introduction}
Correlated electron materials exhibit a variety of behaviors which often arise due a balance between different interactions with similar energy scales. For instance, when a magnetic lattice is immersed in a Fermi sea the RKKY interaction plays an imporant role where the magnetic exchange is no longer limited to the nearest neighbors.~\cite{Ruderman_54, Kasuya_56, Yosida_57} This can lead to many nearly degenerate exchange interactions, and thus the emergence of finely tunable magnetic order.\cite{Selke_84, Selke_88, Bak_82} For some $f$-electron materials (e.g., those containing Ce), the situation is further enriched by the Kondo effect where the conduction electrons hybridize with the $f$-moments, which tends to neutralize the magnetism while enhancing the charge carrier electronic correlations. \cite{Kondo_64, Doniach_77} Together, these factors describe a complicated electronic/magnetic phase space, which has been proposed to host exotic states of matter that go beyond those expected for magnetic insulators or non-frustrated Kondo lattices.\cite{Coleman_10, Lacroix_10} Several promising systems have been investigated so far, including CeSb,\cite{Mignod_77} CeRhIn$_5$,\cite{Das_14} YbAgGe,\cite{Schmiedeshoff_11, Tokiwa_13} Yb$_2$Pt$_2$Pb,\cite{Kim_13} and CePdAl,\cite{Zhao_16, Fritsch_14} but it remains desirable to uncover additional examples.

Meanwhile, the topic of topological protection in strongly correlated electron materials has become a central focus. A quantum state may be called topological when its wavefunctions have some character that remains unchanged by adiabatic deformation of the system: i.e., are specified by a topological invariant.~\cite{Ando_13, Hasan_10} This produces electronic states that are protected against a variety of common mechanisms (e.g., charge carrier scattering due to certain types of chemical/structural disorder) and thus promotes behaviors that are distinct from typical metals and insulators.\cite{Fu_07, Schnyder_08, Hasan_11, Qi_11, Wehling_14} Amongst the $f$-electron materials, the proposed topological insulator SmB$_6$ has been intensely scrutinized,\cite{Xu_13, Jiang_13} and more generally Kondo insulators have been identified as hosts for topologically protected states:\cite{Dzero_10, Alexandrov_13} examples include YbB$_{12}$,\cite{Weng_14} CeRu$_4$Sn$_6$,\cite{Sundermann_15} filled skutterudites,\cite{Yan_12} and CeSb. \cite{Alidoust_16, Guo_16}

Here we report results for the Kondo lattice compound CeSbSe, which crystallizes in the same tetragonal structure as the compounds $WHM$ ($W$ $=$ Zr, Hf, La, $H$ $=$ group IV or V element, and $M$ $=$ group VI element) and LaSbTe, which were recently proposed as potential hosts for topologically protected electronic states.~\cite{Xu_16, Singha_17} Magnetic susceptibility measurements reveal Curie-Weiss behavior, showing that the cerium ions are trivalent at high temperatures. Crystal electric field splitting modifies the low temperature magnetism and complex magnetic ordering with an antiferromagnetic character occurs at $T_{\rm{M}}$ $=$ 3 K, where the magnetization evolves through several fractional integer value steps with increasing magnetic field below 1.8 T: i.e., a possible Devil's staircase is observed.  As evidenced from electrical transport measurements, poor metallic or nearly semiconducting behavior occurs, which likely is due to there being a small electronic density of states and strong scattering of the charge carriers from the $f$-moments. Thus, this material is an intriguing addition to the family of $f$-electron Kondo lattices, and presents the opportunity to study complex magnetism under easily accessed conditions.

\section{Experimental Methods}
Single crystals of CeSbSe were grown using elements with purities Ce (99.9\%), Sb (99.999\%), and Se (99.9999\%) using the iodine vapor transport method. Polycrystalline precursor material was first prepared through direct reaction of the elements at 1000 $^{\circ}$C under vacuum in a sealed quartz tube. The precursor was then sealed with 50 mg of iodine in a quartz tube with dimensions 1.4 cm diameter and 10 cm length. The ampoule was slowly heated until its hot and cold zones were at 950 $^{\circ}$C and 850 $^{\circ}$C, respectively. After soaking under these conditions for 21 days, the ampoule was cooled to room temperature and single crystal specimens were retrieved from the cold zone. Typical crystals were square plates with dimensions up to 0.5 cm width and 0.3 cm thickness. 

CeSbSe was structurally characterized by single crystal x-ray diffraction using an Oxford-Diffraction Xcalibur2 CCD system with graphite monochromated Mo $K\alpha$ radiation. Data was collected using $\omega$ scans with 1$^{\circ}$ frame widths to a resolution of 0.4 \textrm{\AA}, equivalent to $2\theta \approx 125^{\circ}$. Reflections were recorded, indexed and corrected for absorption using the Oxford-Diffraction CrysAlisPro software~\cite{CrysAlisPro}, and subsequent structure determination and refinement was carried out using the single crystal x-ray structure refinement and analysis software package CRYSTALS~\cite{Crystals}, employing Superflip~\cite{Superflip} (programs for solution of crystal structures by charge flipping, to solve the crystal structure). The data quality allowed for an unconstrained full matrix refinement against $F^2$, with anisotropic thermal displacement parameters for all atoms.  A crystallographic information file (CIF) has been deposited with the Inorganic Crystal Structure Database (ICSD CSD No. 432699)~\cite{ICSD}. Electron dispersive spectroscopy chemical analysis measurements were used to confirm the stoichiometry. 

Magnetization $M(H,T)$ measurements were performed for an oriented single crystal at temperatures 1.8 K $<$ $T$ $<$ 300 K under an applied magnetic field of $H$ $=$ 0.5 T and for 0 $<$ $H$ $<$ 7 T at temperatures $T$ $<$ 3.2 K for $H$ applied both parallel $\parallel$ and perpendicular $\perp$ to the $c$-axis using a Quantum Design Magnetic Properties Measurement System 3 in Vibrating Sample Magnetometer mode. The specific heat $C(T)$ was measured for $T$ $=$ 0.8 $-$ 20 K and the electrical resistivity $\rho(T)$ was measured for $T$ = 0.5 K - 300 K using a Quantum Design Physical Property Measurement System (PPMS). Torque magnetometry measurements were performed for $H$ $\leq$ 35 T using a resistive magnet at the National High Magnetic Field Laboratory. Electrical resistivity measurements at pressures up to 1.9 GPa, 1.8 K $<$ $T$ $<$ 300 K, and $H$ $<$ 9 T were performed using a piston cylinder cell that was adapted to the PPMS. 

\section{Results}

\begin{figure}[!t]
    \begin{center}
        \includegraphics[width=0.45\textwidth]{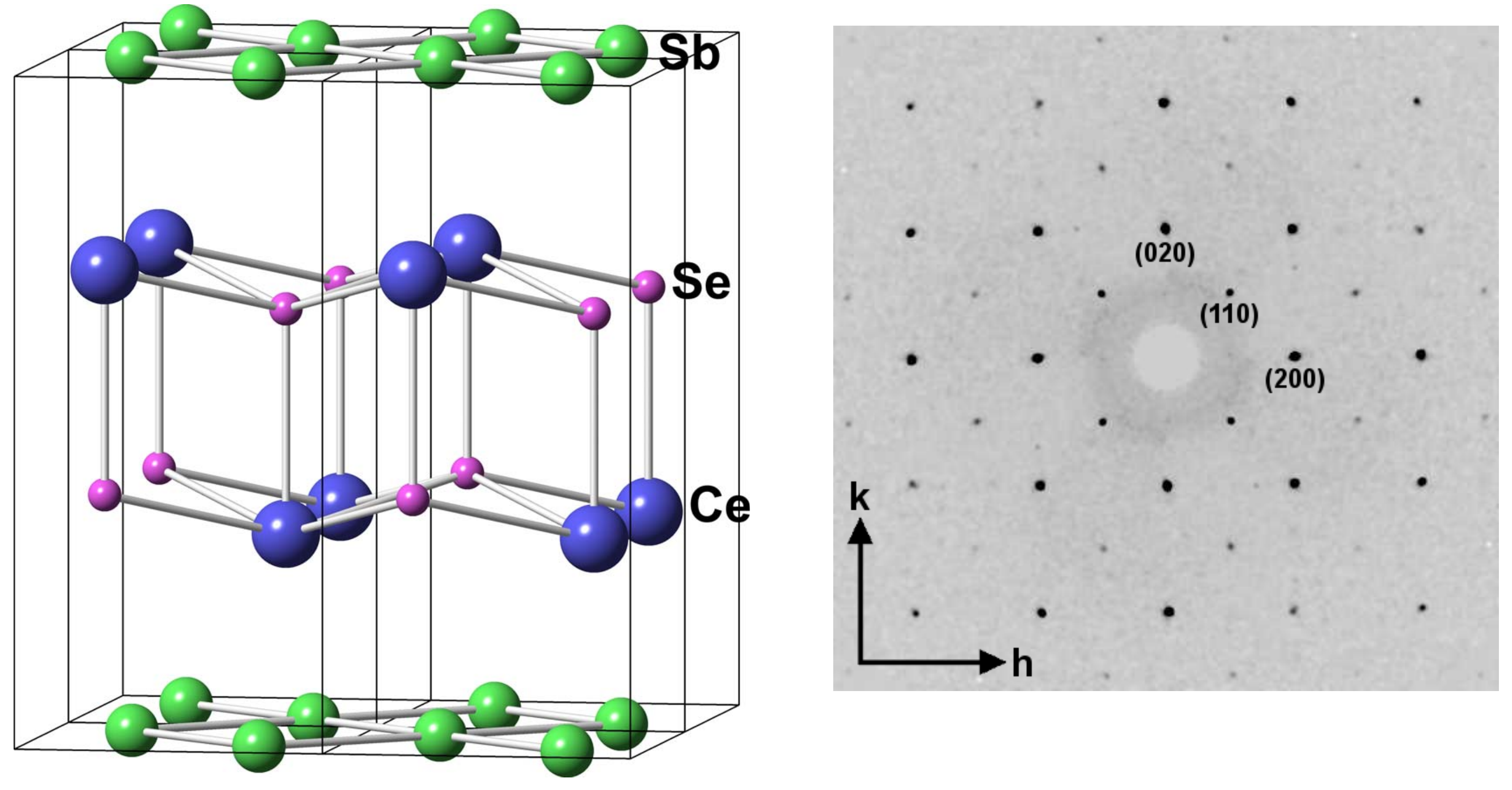}
        \caption{(left) Crystal structure of CeSbSe, which forms in the tetragonal space group $P4/nmm$ (No.129) with unit cell parameters $a=4.2164(1)$~\textrm{\AA} and $c=9.105(1)$~\textrm{\AA}. Details are summarized in Tables~\ref{tbl:xray} and ~\ref{tbl:coord}. (Right) The (h k 0) plane reconstructed from the single crystal x-ray diffraction measurement.}
        \label{fig:struct}
    \end{center}
\end{figure}

CeSbSe crystallizes in the tetragonal space group $P4/nmm$ (No.129) with unit cell parameters summarized in Table~\ref{tbl:xray} (Fig.~\ref{fig:struct} left). This structure can be described as corrugated Ce-Se layers stacked in an ...ABAB... sequence alternated by square-net Sb layers. The Ce ions form a double layer square-net framework in the $ab$-plane where the nearest neighbors are Se ions. This sets the stage for complex magnetic interactions that differ between in-plane nearest, next-nearest, etc. and interplane cerium ions. Single crystal x-ray diffraction measurements reveal that the crystals form as high quality specimens, without evidence for disorder such as site vacancy/interchange or impurities (Fig.~\ref{fig:struct} right). The experimentally generated cif file is included in the supplementary material. Further details of the XRD refinement and the structure are summarized in Tables~\ref{tbl:xray} and ~\ref{tbl:coord}.

\begin{figure}[!t]
    \begin{center}
        \includegraphics[width=0.85\columnwidth]{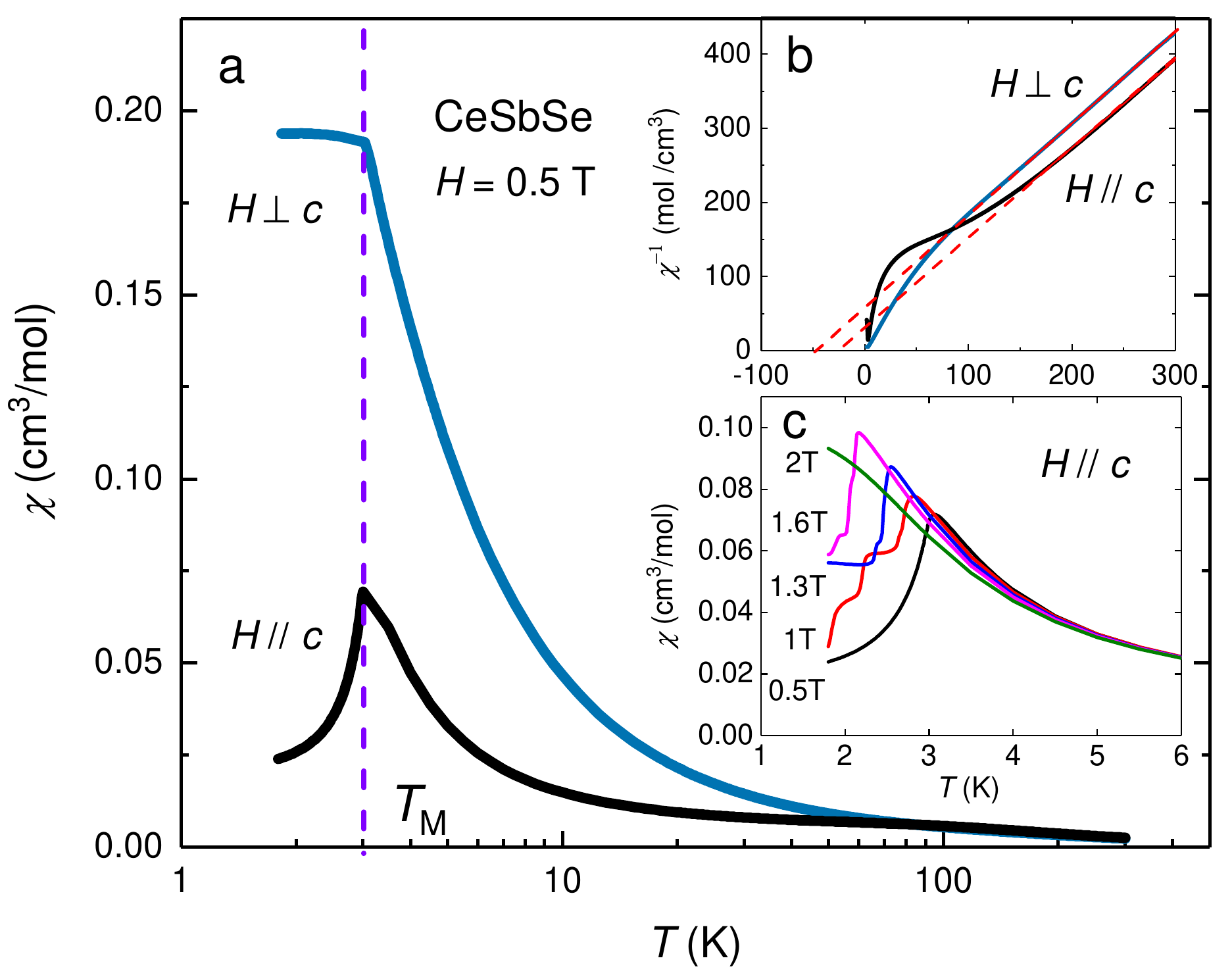}
        \caption{(a) Magnetic susceptibility $\chi$ $=$ $M/H$ vs. temperature $T$ collected in a magnetic field $H$ $=$ 0.5 T applied parallel $\parallel$ and perpendicular $\perp$ to the $c$-axis of CeSbSe. (b) $\chi$$^{-1}$ vs. $T$ for $H$ $\parallel$ and $\perp$ to the $c$-axis. The straight dotted lines are fits to the data using the Curie-Weiss expression, as described in the text. (c) $\chi(T)$ in several magnetic fields $H$ $\leq$ 2 T.} 
        \label{Fig1}
    \end{center}
\end{figure}

The temperature dependent magnetic susceptibility $\chi(T)$ $=$ $M/H$ collected in magnetic fields applied perpendicular $\perp$ and parallel $\parallel$ to the crystallographic $c$-axis are shown in Fig.~\ref{Fig1}a. For 150 K $\leq$ $T$ $\leq$ 300 K, $\chi(T)$ for both directions follows a Curie-Weiss temperature dependence $\chi(T)$ $=$ $C$/($T-\theta$) (Fig.~\ref{Fig1}b), where fits to the data (shown as straight dotted lines in Fig.~\ref{Fig1}b) yield the effective magnetic moments $\mu_{\rm{eff}}$ $=$ 2.54 $\mu$$_{\rm{B}}$ (2.6 $\mu$$_{\rm{B}}$) and Curie-Weiss temperatures $\theta$ $=$ -48.4 K (-31.1 K) for $H$ $\perp$ and $\parallel$ to $c$, respectively. Over the entire temperature range, the $ab$-plane is the easy axis. Crystal electric field splitting of the cerium Hund's rule multiplet causes a deviation from Curie-Weiss behavior for $T$ $\lesssim$ 150 K, and complex magnetic ordering that reduces $\chi$ appears near $T_{\rm{M}}$ $=$ 3 K. As shown in Fig.~\ref{Fig1}c, there is a rapid evolution of $T_{\rm{M}}$ and the appearance of multiple sub-phases for $T$ $<$ $T_{\rm{M}}$ with increasing $H$. This reveals that although the magnetism has an antiferromagnetic character, it is more complex than simple antiferromagnetism and may be frustrated in some way. 
\begin{table}[!b]
    \begin{center}
        \caption[]{Selected single crystal x-ray diffraction data, along with collection and refinement parameters.}
        \begin{tabular}{l l}
            \hline
            \textbf{Compound} & CeSbSe \\
            \hline
            Formula weight (g/mol) & 340.83 \\
            Space group & $P4/nmm$ (\#129)\\
            $a$ (\textrm{\AA}) & 4.2164(1) \\
            $c$ (\textrm{\AA}) & 9.105(1) \\
            Volume (\textrm{\AA}$^{3}$) & 161.87(1) \\
            $Z$ & 2 \\
            Data collection range & $4.48^{\circ}\leq\theta\leq66.53^{\circ}$ \\
            Reflections collected & 4757 \\
            Independent reflections & 907 \\
            Parameters refined & 10 \\
            $R_{1}$, $wR_{2}$ & 0.0660, 0.0679 \\
            Goodness-of-fit on $F^{2}$ & 0.9998 \\
            \hline
        \end{tabular}
        \label{tbl:xray}
    \end{center}
\end{table}

\begin{table}[!b]
   \begin{center}
        \caption[]{Atomic coordinates and equivalent thermal displacement parameters, along with interatomic distances.}
        \begin{tabular}{l l l l l l l}
            \hline
            Atom & Site & $x$ & $y$ & $z$ & $U_{\textrm{eq}}$ (\textrm{\AA}$^{2}$) \\
            \hline
            Ce & 2c & 3/4 & 3/4 & 0.29585(7) & 0.0089(2) \\
            Sb & 2a & 3/4 & 1/4 & 0 & 0.010(2) \\
            Se & 2c & 1/4 & 1/4 & 0.37012(13) & 0.0093(4) \\
            \hline
            \\
            \hline
            \multicolumn{3}{l}{Bond} & \multicolumn{3}{l}{Distance (\textrm{\AA})} \\
            \hline
            \multicolumn{3}{l}{Ce\textemdash Se} & \multicolumn{3}{l}{3.057} (Approx. in $ab$-plane) \\
            \multicolumn{3}{l}{Ce\textemdash Se} & \multicolumn{3}{l}{3.041} (along $c$-axis) \\
            \multicolumn{3}{l}{Sb\textemdash Sb} & \multicolumn{3}{l}{2.981} \\
            \hline
        \end{tabular}
        \label{tbl:coord}
    \end{center}
\end{table}

\begin{figure}[!t]
    \begin{center}
        \includegraphics[width=0.85\columnwidth]{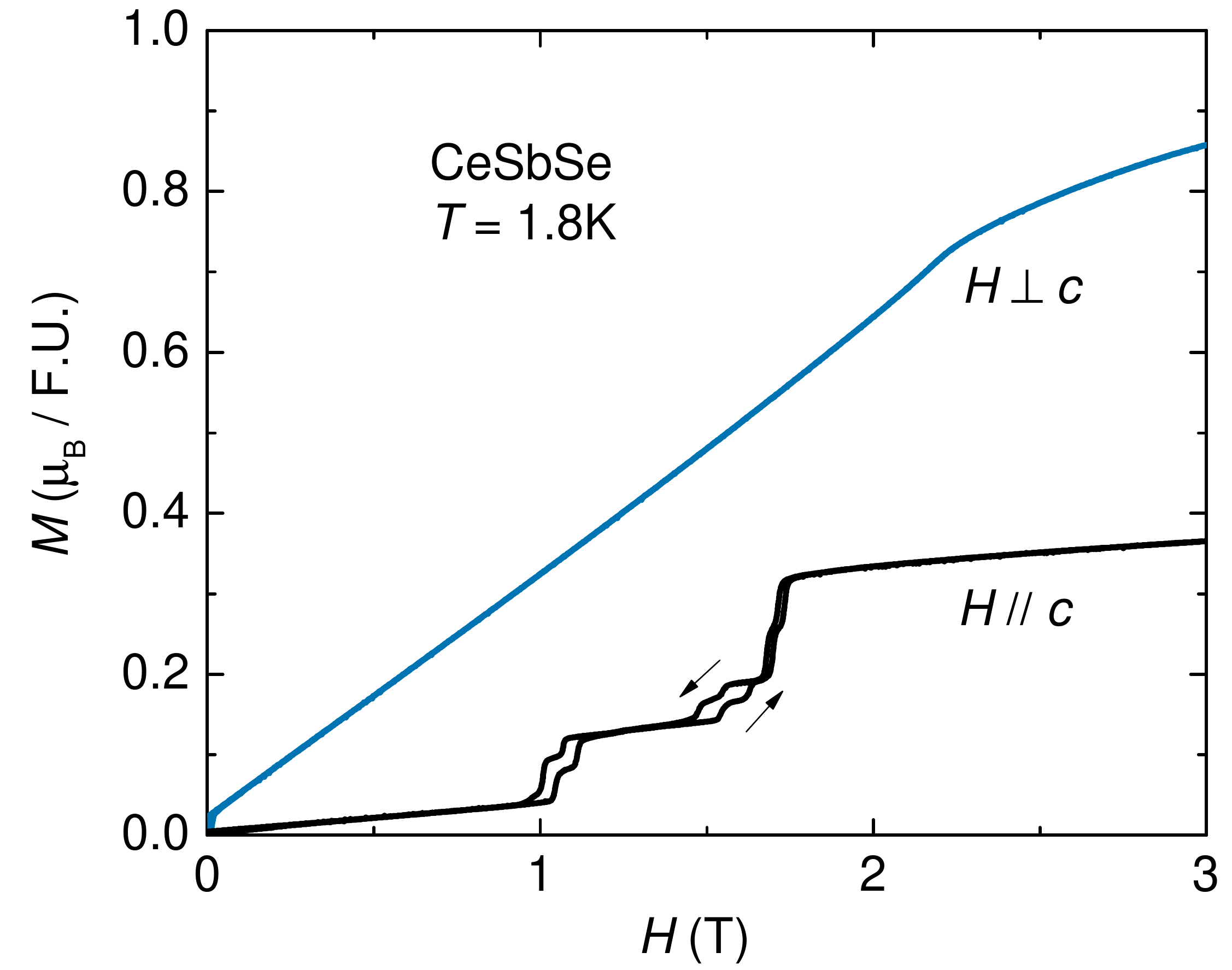}
        \caption{ Magnetization $M$ vs. magnetic field $H$ for $H$ applied parallel $\parallel$ and  perpendicular $\perp$to the $c$-axis of CeSbSe.} 
        \label{Fig3}
    \end{center}
\end{figure}

\begin{figure}[!t]
    \begin{center}
        \includegraphics[width=0.85\columnwidth]{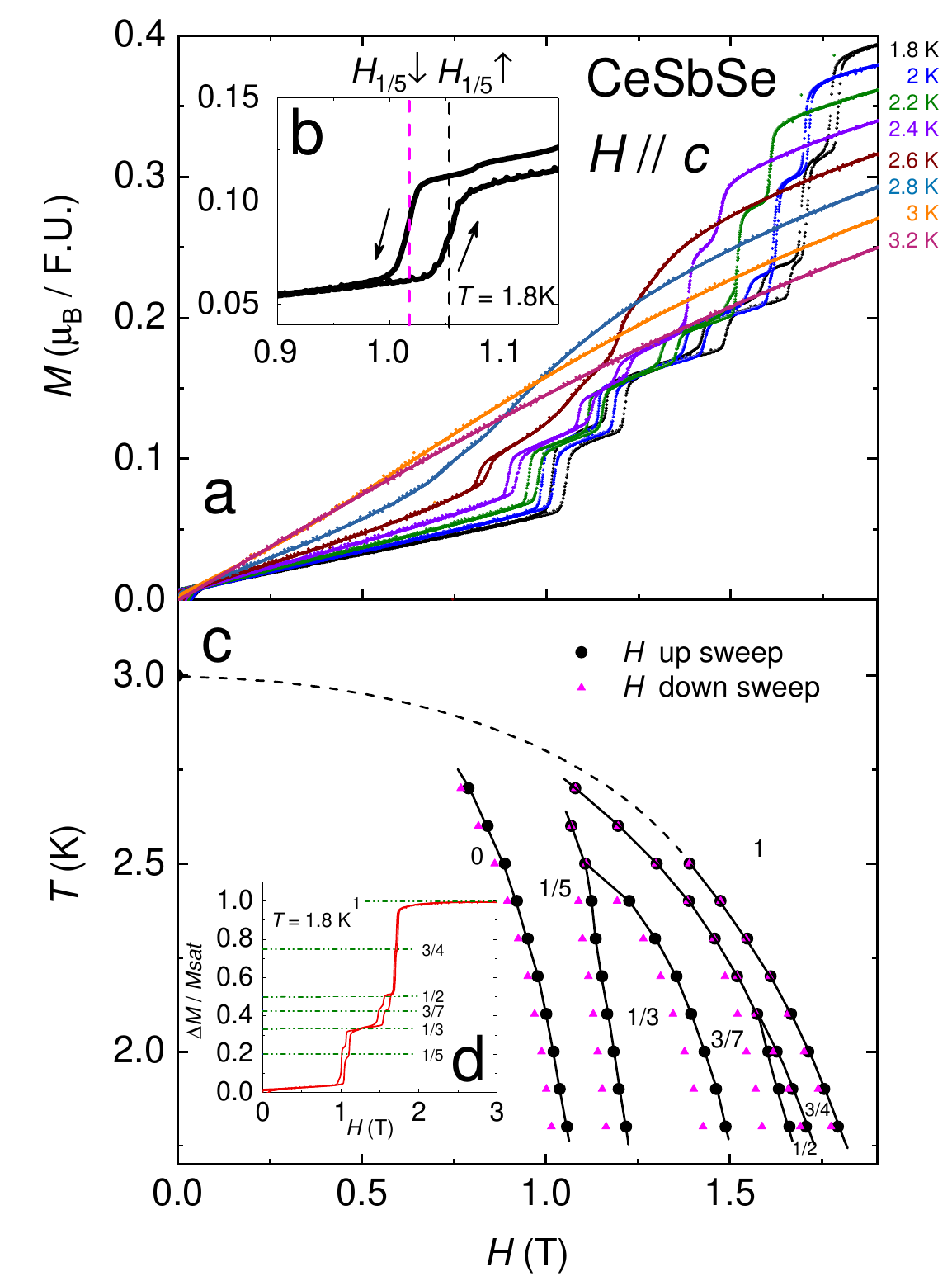}
        \caption{(a) Magnetization $M$ vs magnetic field $H$ for $H$ applied parallel $\parallel$ to the $c$-axis. Several hysteretic phase transitions occur, which broaden and decrease in $H$ with increasing $T$. (b) Zoom of $M(H)$ in the region of the first hysteretic phase transition. Up-sweep and down-sweep directions are indicated by arrows. (c) $T-H$ phase diagram for $H$ $\parallel$ $c$ constructed from the $M(H)$ data in panel a. (d) The magnetization $M$ where the low field linear slope region has been subtracted from the data, and afterwards normalized to the value of $\Delta$$M$ at 2.5 T. The data are plotted this way to emphasize the relative sizes of the steps, which all assume integer fractional values.} 
        \label{Fig4}
    \end{center}
\end{figure}

In order to probe the magnetic order, we performed magnetization measurements for $T$ $\leq$ $T_{\rm{M}}$ (Figs.~\ref{Fig3} and ~\ref{Fig4}) where we find that $M(H)$ for $H$ $\perp$ $c$ evolves monotonically and exhibits a weak kink near $H$ $=$ 2.2 T for $T$ $=$ 1.8 K. In contrast, for $H$ $\parallel$ $c$ there is a complex evolution of the magnetic state for $H$ $\lesssim$ 1.8 T and $T$ $\leq$ $T_{\rm{M}}$. $M(H)$ is linear starting from $H$ $=$ 0, as would be expected in an antiferromagnetically ordered state. At the lowest temperatures, there is an abrupt hysteretic increase in $M$ near $H$ $=$ 1 - 1.3 T that features several sub-steps. This is followed by another linear increase region, a hysteretic several-step region for $H$ $=$ 1.3 - 1.68 T, a sharp step region for $H$ $=$ 1.68 - 1.85 T, and finally a weakly saturating region for $H$ $>$ 1.85 T. After subtracting a linear background, it is evident that the steps in the magnetization have fractional integer values (Fig.~\ref{Fig4}d). To search for additional steps in the magnetization, torque magnetometry measurements were performed for $H$ $\leq$ 35 T (not shown) but no additional sharp features were detected. The complex steps in the magnetization evolve strongly with temperature and disappear for $T$ $>$ $T_{\rm{M}}$ (Fig.~\ref{Fig4}a). 

From the features identified in $M(H)$ for $H$ $\parallel$ $c$ we construct the $T-H$ phase diagram shown in Fig.~\ref{Fig4}c. At base temperature and starting from $H$ $=$ 1.05 T, we observe six distinct regions, all of which are entered through a hysteretic phase transition. For $H$ $\geq$ 1.85 T, the magnetization reaches a saturation value near 0.4 $\mu$$_{\rm{B}}$.  As temperature increases, several of these phases collapse or merge into each other as the features that are used to define them broaden in $H$. This phase diagram is reminiscent of that seen for other ``Devil's staircase" materials such as CeSb,\cite{Mignod_77} CeRh$_3$Si$_2$\cite{Pikul_10} and TbMnO$_3$. \cite{Kimura_05} It is noteworthy that this phase diagram is arborescent, and features many first order lines and phase boundaries that terminate at finite temperatures. 

\begin{figure}[!t]
    \begin{center}
        \includegraphics[width=0.85\columnwidth]{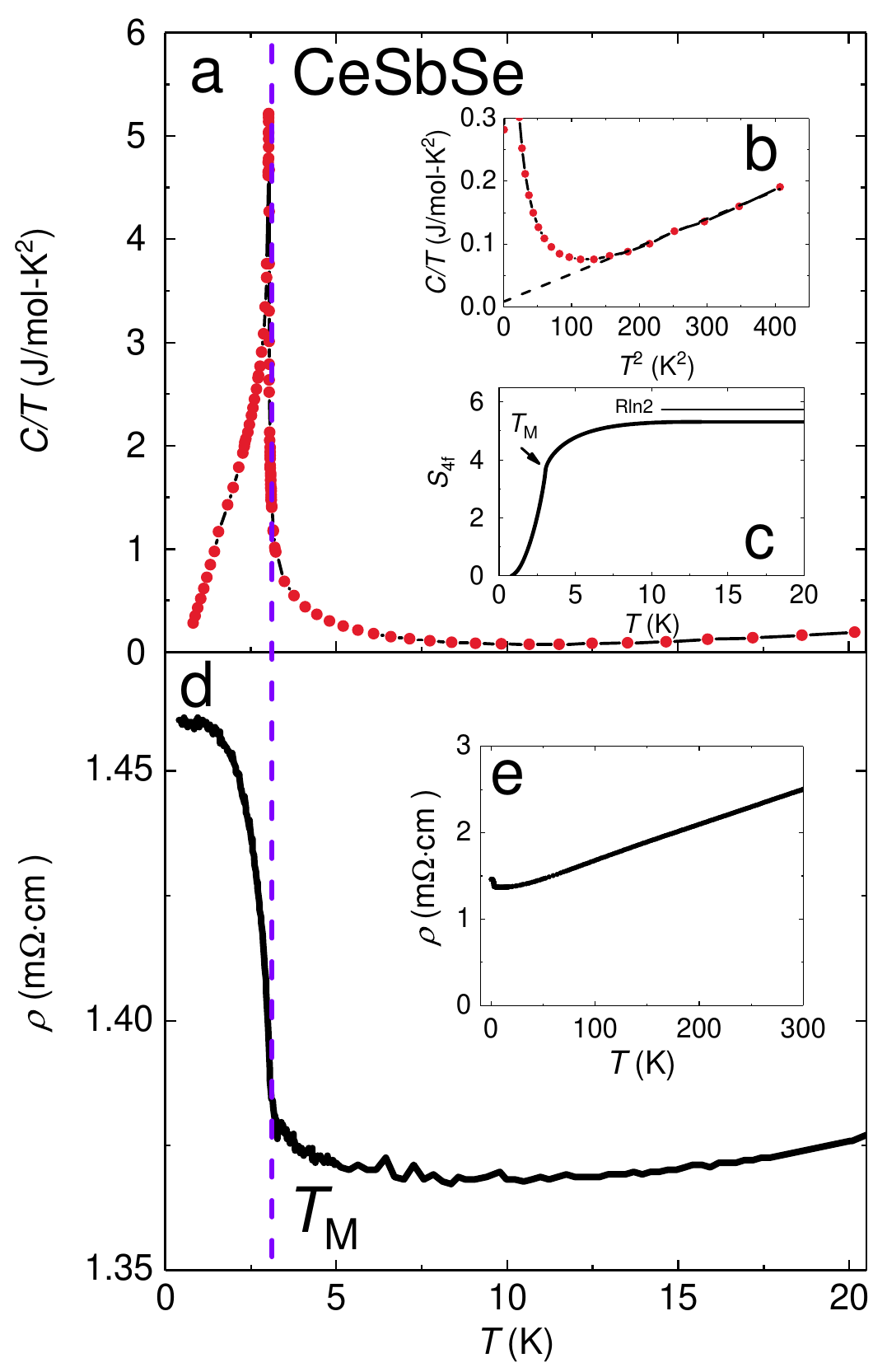}
        \caption{(a) Heat capacity divided by temperature $C/T$ vs $T$ for CeSbSe. (b) $C/T$ vs. $T^2$. The dotted line is a fit to the data using the expression for a Fermi liquid $C/T$ $=$ $\gamma$ $+$ $\beta$$T^2$, where $\gamma$ is the electronic coefficient of the heat capacity and $\beta$$T^2$ is the low temperature limit of the lattice term. (c) The 4$f$ contribution to the entropy $S_{4f}$ vs. $T$, calculated as described in the text. (d) Electrical resistivity $\rho$ vs. $T$ for $T$ $\leq$ 20 K. (e) $\rho(T)$ over the temperature range 500 mK $\leq$ $T$ $\leq$ 300 K.} 
        \label{Fig2}
    \end{center}
\end{figure}

The heat capacity divided by temperature $C/T$ vs. $T$ data are summarized in Fig.~\ref{Fig2}a. The evolution of $C/T$ for $T$ $\geq$ $T_{\rm{M}}$ (Fig.~\ref{Fig2}b) is consistent with the Fermi liquid expression $C/T$ $=$ $\gamma$ $+$ $\beta$$T^2$, where $\gamma$ $=$ 8.6 mJ/mol-K$^2$ is the electronic coefficient of the heat capacity and $\beta$ $=$ 0.44 mJ/mol-K$^4$ is the coefficient of the low temperature limit of the lattice term. From $\beta$, we calculate a Debye temperature $\theta$$_{\rm{D}}$ = 237 K, but emphasize that this is only a rough estimate since our calculation does not explicitly account for other factors: e.g., the influence of crystal electric field splitting. There is a sharp and nearly lambda-like anomaly at $T_{\rm{M}}$ $=$ 3 K, which suggests both that there is limited structural/chemical disorder and that magnetic fluctuations persist above the ordering temperature. In Fig. ~\ref{Fig2}c we show the 4$f$ contribution to the entropy $S_{\rm{4f}}$ vs. $T$ obtained by subtracting the $\beta$$T^2$ term and integrating $C/T$ from 0.8 K. This quantity underestimates $S_{\rm{4f}}$ because the lower limit of integration only extends to 800 mK, but nonetheless it provides some useful insight. While $S_{\rm{4f}}$ increases abruptly at $T_{\rm{M}}$ = 3 K, it only reaches 66\% of $R\ln2$ at this temperature and recovers 90 \% of $R\ln2$ near $T=10$ K.  Such behavior is typical for many Ce-based materials, and may provide evidence that the cerium magnetic moment is partly compensated by the conduction electrons through the Kondo interaction.\cite{Hegger_00}

In Figs.~\ref{Fig2}d,e we show the temperature dependence of the electrical restivity $\rho$. Starting from $T$ $=$ 300 K, $\rho$ decreases slowly with decreasing $T$. This is distinct both from the behavior of a simple metal or a metallic strongly correlated Kondo lattice and indicates a small charge carrier density along with electronic scattering from the $f$-moments. In agreement with $\chi(T)$ and $C/T$, $\rho(T)$ shows a sharp feature at $T_{\rm{M}}$, where $\rho$ abruptly increases. This differs from what is seen in many metallic magnets where the removal of spin scattering results in a decrease of $\rho$ at the ordering temperature. Instead, it is consistent with the opening of a gap in the electronic spectrum, and the partial loss of carriers. At low temperature, $\rho(T)$ finally saturates towards a large residual resistivity $\rho$$_0$ $=$ 1.26 m$\Omega$ cm. 

\begin{figure}[!tht]
    \begin{center}
        \includegraphics[width=0.85\columnwidth]{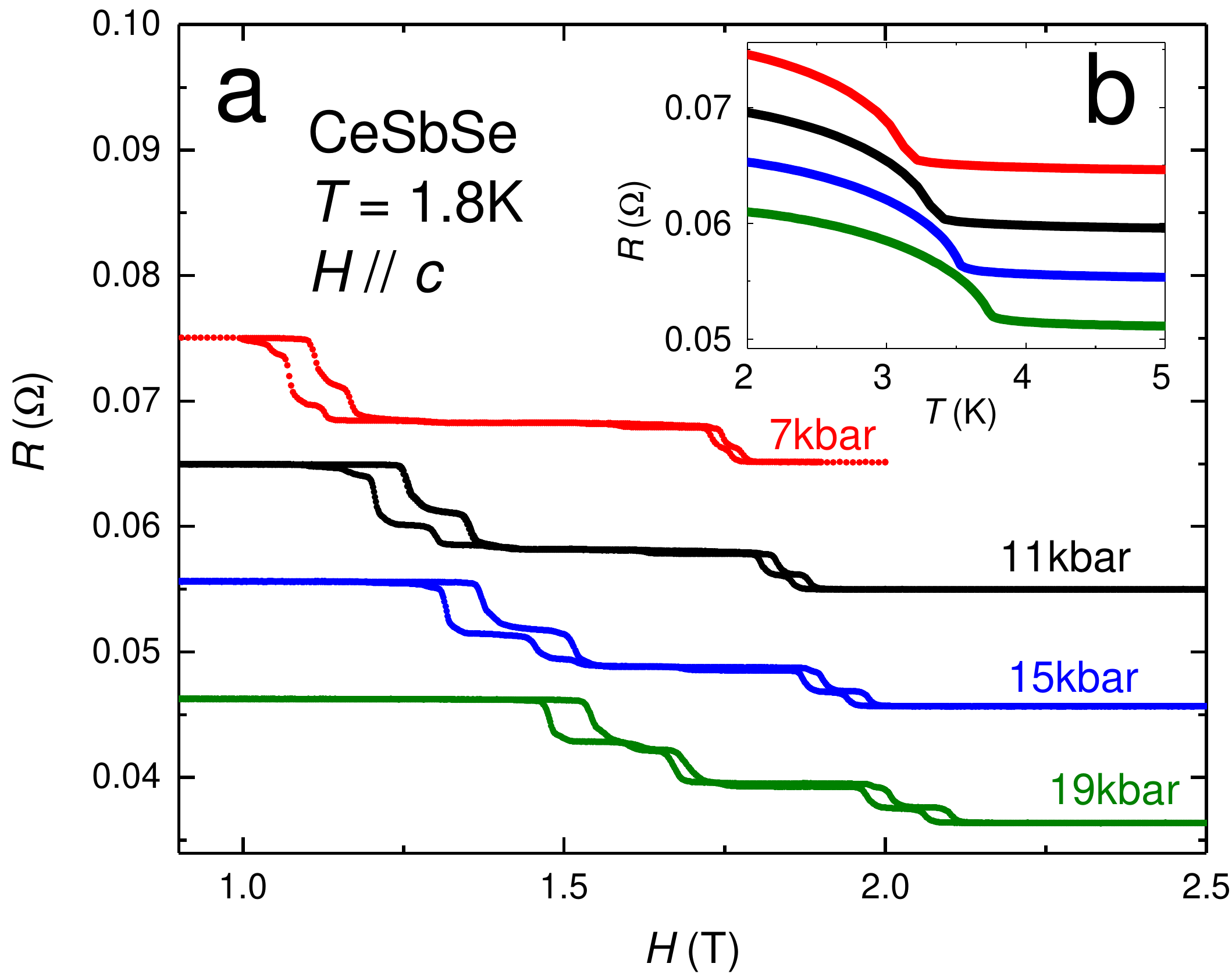}
        \caption{(a) Electrical resistance $R$ vs. magnetic field $H$ with the $H$ and the electrical current applied along the $c$-axis for hydrostatic pressures $P$ $\leq$ 19 kbar in CeSbSe. (b) $R(T)$ near the magnetic phase transition at different $P$.} 
        \label{Fig6}
    \end{center}
\end{figure}

We additionally studied the electrical resistivity under applied pressure, where transitions associated with the Devil's staircase appear as sharp reductions in resistance (Fig.~\ref{Fig6}). We find that (1) $T_{\rm{M}}$ increases with $P$ and (2) the general features of the field induced phases are preserved but move to higher $H$. This can be understood by considering that the magnetic exchange strength for the RKKY interaction is given by the expression $j$ $=$ $j_0$$^2$$N(E_F)$cos(2$k_F$$r$)/$|$$k_Fr$$|$$^3$, where $r$ is the distance between localized magnetic moments, $N(E_F)$ is the density of states at the Fermi energy, and $k_F$ is the Fermi wave vector. A decreasing $r$ results in a strengthening $j$ and, in the regime where the RKKY interaction dominates over the Kondo interaction, an increasing ordering temperature.~\cite{Doniach_77}

\section{Discussion}
We have shown that CeSbSe exhibits complex magnetism featuring a possible Devil's staircase. This behavior occurs in the context of an $f$-electron lattice with many characteristics that are typical for intermetallics with trivalent cerium. At high temperatures the $f$-electrons are well localized, as evidenced by the Curie-Weiss behavior of the magnetic susceptibility. Crystal electric field splitting of the Hund's rule multiplet results in a deviation from Curie Weiss behavior for $T$ $\lesssim$ 150 K and a low temperature magnetic saturation moment in applied magnetic field that is smaller than the value of the full $J$ $=$ 5/2 multiplet. There is evidence for hybridization between the $f$- and conduction electrons, including that the magnetic entropy reaches $\approx$ 0.66Rln2 at $T_{\rm{M}}$. This value is reduced from expectations for a doublet ground state, likely because the magnetism is compensated through the Kondo interaction. The pressure dependence of $T_{\rm{M}}$ suggests that the RKKY interaction dominates over the Kondo interaction at low temperature.~\cite{Doniach_77}

The high residual resistivity (and weak evolution in $\rho(T)$) indicates that there is a low density of charge carriers which are strongly scattered from the local $f$-moments. The nearly linear behavior observed in the resistivity for temperatures ranging from 30 to 300 K already is evidence for an anomalous scattering mechanism that may involve a Bosonic mode. We emphasize that this behavior likely is not due to disorder scattering because there is ample evidence that the crystals are highly ordered; (1) single crystal x-ray diffraction measurements show that there is little site vacancy or interchange, (2) the magnetic phase transitions are sharp both in temperature and magnetic field, and (3) earlier studies of the isolelectronic analogues CeSbTe and LaSbTe support the view that the semimetallic behavior results from the $f$-electron Kondo lattice: while CeSbTe exhibits poor metallic or doped semiconducting behavior,\cite{Wang_01} its non-4$f$ analogue LaSbTe has a small residual resistivity.\cite{Singha_17} A low charge carrier density also accounts for the small $\gamma$ and may still imply that the charge carrier effective mass is enhanced through the Kondo interaction.

From these results, the picture that emerges is that CeSbSe is a low charge carrier density semimetal where there is mild `frustration' due to the Kondo interaction partly screening the local moments, which mainly interact through the RKKY interaction. In this context, we suggest that the magnetic complexity occurs mainly due to a balance between nearest-, next nearest-, etc. neighbor interactions. The low charge carrier density likely plays an important role because the RKKY interaction strength depends both on $N(E_F)$ and $k_{\rm{F}}$. A well known model that considers these factors is the anisotropic next nearest neighbor Ising (ANNNI) model.\cite{Selke_84, Selke_88, Bak_82} While this picture has been discussed for other non-geometrically frustrated materials it is not yet clear whether it, or one of its variants, apply to CeSbSe. Measurements such as neutron scattering to measure the magnetic field dependence of the magnetic ordering wave vector are needed to clarify this question.


We finally point out that it remains to be determined whether CeSbSe hosts topologically protected electronic states, similar to the related compounds $WHM$ ($W$ $=$ Zr, Hf, La, $H$ $=$ group IV or V element, and $M$ $=$ group VI element) and LaSbTe.\cite{Singha_17}It is attractive to consider this scenario because it would result in a situation where the topological state is coupled to complex magnetic order: e.g., as has been explored for CeSb.~\cite{Alidoust_16} Before progress can be made to clarify the electronic state, it will be necessary to better understand the band structure: e.g., through electronic structure calculations and measurements such as ARPES to probe the fermiology.

\section{Summary}
Results are presented for CeSbSe, which exhibits complex magnetic ordering at $T_{\rm{M}}$ $=$ 3 K. The application of a magnetic field results in a cascade of magnetically ordered states for $H$ $\lesssim$ 1.8 T which are characterized by fractional integer size steps: a possible Devil's staircase is observed. These behaviors are considered in the context of a Kondo lattice with a small charge carrier density where the $f$-moments are incompletely screened, resulting in a fine balanced magnetic interaction between different Ce neighbors that is mediated by the RKKY interaction.

\section{Acknowledgements}
This work was performed at the National High Magnetic Field Laboratory (NHMFL), which is supported by National Science Foundation Cooperative Agreement No. DMR-1157490, the State of Florida and the DOE. A portion of this work was supported by the  NHMFL User Collaboration Grant Program (UCGP). RB, DG, and TEAS were supported as part of the Center for Actinide Science and Technology (CAST), an Energy Frontier Research Center funded by the U.S. Department of Energy, Office of Science, Basic Energy Sciences under Award Number DE-SC0016568. LB is supported by DOE-BES through award DE-SC0002613.

\end{document}